\documentclass[aps,prd,twocolumn,showpacs,nofootinbib]{revtex4-2}

\usepackage{graphicx}
\usepackage{amsmath}
\usepackage{amssymb}
\usepackage{bm}
\usepackage{bbm}
\usepackage{color}
\usepackage{slashed}
\usepackage[colorlinks=true,linkcolor=blue,citecolor=blue, urlcolor=blue]{hyperref}

\begin{document}

\title{Next-to-leading order QCD and relativistic corrections to $Z \to J/\psi+\Upsilon(nS)$}

\author{Guang-Yu Wang$^{a}$}
\email{guangyuwang@sjtu.edu.cn}
\author{Xu-Chang Zheng$^{b}$}
\email{zhengxc@cqu.edu.cn}
\author{Guang-Zhi Xu$^{c}$}
\email{xuguangzhi@lnu.edu.cn}

\affiliation{$^a$Shanghai Key Laboratory for Particle Physics and Cosmology,
Key Laboratory for Particle Astrophysics and Cosmology (MOE),
School of Physics and Astronomy, Shanghai Jiao Tong University, Shanghai 200240, China\\
$^b$ Department of Physics, Chongqing Key Laboratory for Strongly Coupled Physics, Chongqing University, Chongqing 401331, People's Republic of China \\
$^c$ Department of Physics, Liaoning University, Shenyang 110036, People's Republic of China}

\begin{abstract}
In this paper, we calculate the decay widths and branching fractions for the decays $Z \to J/\psi+\Upsilon(nS)$ ($n=1,2,3$) at future super $Z$ factory and at the CEPC/FCC-ee, including both the relativistic and QCD corrections within the framework of nonrelativistic QCD. Both the relativistic and QCD corrections are found to be large and negative. Compared to the leading-order results, the decay widths are significantly reduced by the higher-order corrections due to significant numerical cancellations. Despite the resulting large theoretical uncertainties, sizable event rates from these rare decay channels could still be anticipated at future high-luminosity electron-positron colliders running around the Z-pole. Ultimately, our results provide an essential theoretical baseline, highlighting the necessity of incorporating even higher-order corrections and resummation techniques for future precision phenomenological studies.
\end{abstract}

\maketitle

\section{Introduction}
\label{secIntro}

As a fundamental electroweak gauge boson in the Standard Model (SM), the $Z$-boson's properties offer a rigorous benchmark for testing perturbative QCD (pQCD) and nonperturbative heavy quark dynamics \cite{Dong:2022ayy}. Among various $Z$-boson decay channels, the exclusive associated production of $J/\psi$ and $\Upsilon(nS)$ ($n=1,2,3$) serves as an ideal laboratory. Being an exclusive process, it avoids theoretical complexities such as fragmentation and multi-parton scattering inherent in inclusive production \cite{Bergstrom:1990bu}. Furthermore, this mixed-flavor channel ($c\bar{c} + b\bar{b}$) facilitates a clean test of heavy quarkonium dynamics across different energy scales within a single process.

The theoretical description of such processes relies on the nonrelativistic QCD (NRQCD) factorization framework \cite{Bodwin:1994jh}. Within NRQCD, decay rates are expressed as products of perturbatively calculable short-distance coefficients (SDCs) and universal nonperturbative long-distance matrix elements (LDMEs) \cite{Bodwin:1996tg}. This systematic separation addresses the hierarchical scales of heavy quarkonium by organizing contributions into a double expansion in the strong coupling constant $\alpha_s$ and the heavy quark relative velocity $v$ \cite{Lepage:1992tx, Chen:2021tmf}. Both expansions are essential for precision, and their validity has been demonstrated across diverse channels from $e^+e^-$ annihilation to hadron collisions \cite{Xu:2012am, He:2014sga, Zhang:2005cha, Gong:2007db, Zhang:2008gp, Brambilla:2010cs, Dong:2011fb, Sun:2018rgx, Sun:2021tma}.

Experimental searches have already provided constraints on related same-flavor channels. The CMS Collaboration at the LHC investigated $Z\to J/\psi+J/\psi$ and $Z\to \Upsilon\Upsilon$ decays, establishing upper limits such as ${\rm Br}_{Z\to J/\psi J/\psi}<1.1\times 10^{-6}$ and ${\rm Br}_{Z\to\Upsilon(1S)\Upsilon(1S)} \leq 1.8\times 10^{-6}$ \cite{CMS:2019wch, CMS:2022fsq}. On the theoretical side, next-to-leading order (NLO) QCD corrections for these same-flavor decays have been calculated, revealing that higher-order effects significantly reduce the decay rates \cite{Li:2023tzx, Luo:2022ugd, Li:2024zun}. Recently, the combined impact of NLO and relativistic corrections for $Z \to J/\psi+J/\psi$ was also investigated \cite{Wang:2024chh}.

However, despite these advancements for same-flavor final states, a complete theoretical treatment for the mixed-flavor channels $Z \to J/\psi+\Upsilon(nS)$ that incorporates both NLO QCD and relativistic corrections is still missing. In recent years, several high-luminosity $e^+e^-$ colliders have been proposed, such as the International Linear Collider (ILC) \cite{ILC:2013jhg}, the Circular Electron Positron Collider (CEPC) \cite{CEPCStudyGroup:2018ghi}, the FCC-ee \cite{FCC:2018evy}, and the Super Z factory \cite{zfactory}. Given the enormous number of Z-boson events expected at these future facilities, precise theoretical predictions for these mixed-flavor channels are indispensable.

In this paper, we provide a complete analysis of $Z \to J/\psi+\Upsilon(nS)$ ($n=1,2,3$)  decays up to $\mathcal{O}(\alpha_s)$ and $\mathcal{O}(v^2)$ accuracy. Our study accounts for the significant cancellations between the leading-order (LO) contribution and the negative $\mathcal{O}(\alpha_s)$ and $\mathcal{O}(v^2)$ corrections, and provides updated branching fractions with a rigorous treatment of theoretical uncertainties. Our results show that both QCD and relativistic corrections are sizable and must be included for reliable predictions of these processes.

The organization of this paper is as follows. Section \ref{sec2} establishes the NRQCD factorization framework. Section \ref{sec3} outlines the computational methodology for pQCD calculations. In Section \ref{sec4}, we present the numerical results and discuss their physical implications. We conclude in Section \ref{sec5}.

\section{NRQCD factorization formalism for the decay widths}
\label{sec2}
In the NRQCD factorization approach, the color-singlet amplitude for $Z \to J/\psi+\Upsilon(nS)$, including corrections up to ${\cal O}(\alpha_{s})$ and ${\cal O}(v^2)$, is given by
\begin{eqnarray}
&&{\cal M}_{Z \to H_{_1} + H_{_2}}\nonumber\\
&&=\sqrt{4m_{H{_1}}m_{H{_2}}}\left(c_0^{(0)}+c_0^{(1)}\frac{\alpha_{s}}{\pi}+c_{_2,_1}^{(0)} \langle v^2 \rangle_{H_{_1}} +c_{_2,_2}^{(0)} \langle v^2 \rangle_{H_{_2}}\right)\nonumber\\
&&\times
\langle H_{_1} \vert \psi_1^\dagger {{\bm \sigma}\cdot {\bm \epsilon}} \chi_1 \vert 0  \rangle
\langle H_{_2} \vert \psi_2^\dagger {{\bm \sigma}\cdot {\bm \epsilon}} \chi_2 \vert 0 \rangle,
\label{eq.nrqcdfact}
\end{eqnarray}
where $H_1$ ($J/\psi$) and $H_2$ ($\Upsilon(nS)$) represent the final-state quarkonium states. The coefficients c represent the SDCs, where $c_0^{(0)}$ is the leading-order (LO) contribution, $c_0^{(1)}$ denotes the next-to-leading-order (NLO) QCD correction at 
${\cal O}(\alpha_{s})$, and $c_{2,1}^{(0)}$ and $c_{2,2}^{(0)}$ incorporate the relativistic corrections at ${\cal O}(v^2)$ for $H_1$ and $H_2$, respectively. The LDMEs $\langle H_i \vert \psi_i^\dagger {{\bm \sigma}\cdot {\bm \epsilon}} \chi_i \vert 0  \rangle$ (for $i=1,2$) describes the non-perturbative production of the vector quarkonium states $J/\psi$ and $\Upsilon(nS)$. Here, the Pauli spinor fields $\psi_i$ and $\chi_i$ describe the creation of the heavy quark $Q_{i}$ and the annihilation of the antiquark $\bar{Q}_{i}$, respectively, where the flavor index $i$ corresponds to the charm quark ($Q_1 = c$) or bottom quark ($Q_2 = b$). The dimensionless quantity $\langle v^2 \rangle_{H_i}$, defined as 
\begin{equation}
\langle v^2 \rangle_{H_{_i}}=\frac{\langle H_{_i} \vert \psi_i^\dagger \left(-\frac{i}{2}\tensor{\bf D}\right)^2{{\bm \sigma}\cdot {\bm \epsilon}}  \chi_i \vert 0 \rangle}{m_{Q_i}^2 \langle H_i \vert \psi_i^\dagger {{\bm \sigma}\cdot {\bm \epsilon}}  \chi_i \vert 0 \rangle},
\end{equation}
characterizes the ${\cal O}(v^2)$ corrections for $J/\psi$ and $\Upsilon(nS)$, where the covariant derivative operator is given by $\psi_i^\dagger \tensor{\bf D}\chi_i\equiv \psi_i^\dagger {\bf D}\chi_i-({\bf D} \psi_i)^\dagger \chi_i$. The factor $\sqrt{4m_{H_1}m_{H_2}}$ arises from that we use relativistic normalization for
the quarkonium states in ${\cal M}_{Z \to H_{_1} + H_{_2}}$, but we use conventional nonrelativistic normalization for the NRQCD matrix
element on the right-hand side of Eq. (\ref{eq.nrqcdfact}). It can be expressed in terms of heavy quark mass and $\langle v^{2}\rangle_{H_{i}}$ using the Gremm-Kapustin relation \cite{Gremm:1997dq}, i.e.,
\begin{equation}
m_{H_{i}}^{2} \approx 4 m_{Q_{i}}^{2}\left(1+\left\langle v^{2}\right\rangle_{H_{i}}\right).
\end{equation}

Having the amplitude, the decay width $\Gamma_{Z \to H_{_1} + H_{_2}}$ can be written as
\begin{equation}
\Gamma_{Z \to H_{_1} + H_{_2}}=\frac{1}{3} \frac{\vert \bm{p}  \vert}{8\pi\, m_{_Z}} \sum \vert {\cal M}_{Z \to H_{_1} + H_{_2}} \vert^2,
\label{eq.gamma}
\end{equation}
where  $\sum$ denotes the sum over the polarizations of the initial-state Z boson and the two final-state quarkonium mesons, the factor $1/3$ accounts for the average over the three polarization states of the Z boson, $\bm{p}$ represents the magnitude of the three-momentum of either final-state quarkonium $H_i$ (for $i=1,2$) in the rest frame of the Z boson, and the magnitude of $\bm{p}$ is given explicitly by 
\begin{equation}
\vert \bm{p}  \vert=\frac{\lambda^{1 / 2}\left(m_{_Z}^{2}, m_{H_{_1}}^{2}, m_{H_{_2}}^{2}\right)}{2 m_{_Z}},
\end{equation}
where the Källén function $\lambda(x, y, z)$ is defined as  $\lambda(x, y, z) \equiv x^{2}+y^{2}+z^{2}-2(x y+y z+x z)$.

\section{Details for the perturbative QCD calculation}
\label{sec3}
In the NRQCD framework, the SDCs are derived by matching the pQCD amplitude to the corresponding amplitude in the NRQCD effective theory for the production of on-shell heavy quark-antiquark pairs. Focusing on the production of intermediate $Q_i\bar{Q}_i$ pairs, our calculation is confined to the spin-color configurations $n_1=\,^3S_1^{[1]}$ and $n_2=\,^3S_1^{[1]}$, which correspond to the intermediate states leading to the final $J/\psi$ and $\Upsilon(nS)$ mesons, respectively. The pQCD amplitude for the decay $Z \to Q_1\bar{Q}_1[\,^3S_1^{[1]}] + Q_2\bar{Q}_2[\,^3S_1^{[1]}]$ can be factorized as
\begin{eqnarray}
&&{\cal M}_{Z \to Q_1\bar{Q}_1[\,^3S_1^{[1]}] + Q_2\bar{Q}_2[\,^3S_1^{[1]}]}\nonumber\\
=&&\left(c_0^{(0)}+c_0^{(1)}\frac{\alpha_{s}}{\pi}+c_{_2,_1}^{(0)} v^2_{_1} +c_{_2,_2}^{(0)} v^2_{_2}\right)
\nonumber\\
\times&&\langle Q_1\bar{Q}_1 \vert \psi_1^\dagger {{\bm \sigma}\cdot {\bm \epsilon}} \chi_1  \vert 0 \rangle\langle Q_2\bar{Q}_2 \vert \psi_2^\dagger {{\bm \sigma}\cdot {\bm \epsilon}} \chi_2  \vert 0 \rangle.
\label{eq.nrqcdfact-QQbar}
\end{eqnarray}

\begin{figure}[htbp]
\includegraphics[width=0.45\textwidth]{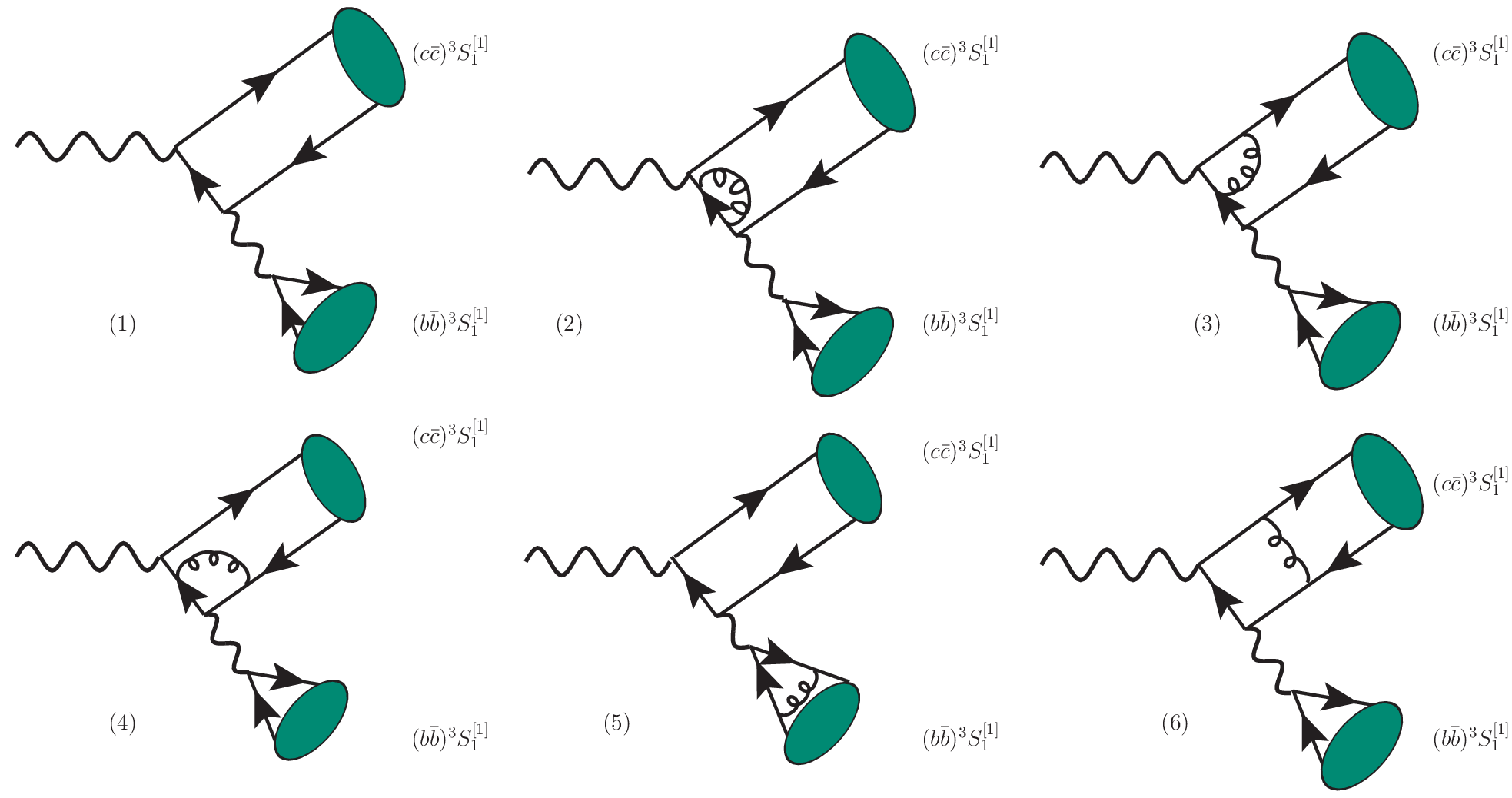}
\caption{Representative typical Feynman diagrams for $Z \to Q_{1}\bar{Q_{1}}[^3S_1^{[1]}]+Q_{2}\bar{Q_{2}}[^3S_1^{[1]}]$. Diagram (1) is the tree-level diagram. Diagrams (2)-(6) depict the NLO QCD corrections to (1).
 } \label{feydiagram}
\end{figure}

For the decay $Z \to Q_1\bar{Q}_1[\,^3S_1^{[1]}] + Q_2\bar{Q}_2[\,^3S_1^{[1]}]$, there are four tree-level Feynman diagrams and twenty one-loop Feynman diagrams in total. One tree-level diagram is shown in Fig.\ref{feydiagram}(1), and five one-loop diagrams are displayed in Figs.\ref{feydiagram}(2)-(6). The remaining tree-level and one-loop diagrams can be obtained from those in Fig.\ref{feydiagram} by exchanging the flavors of the quark-antiquark pairs or reversing the fermion lines, due to flavor and parity symmetry.

In an arbitrary frame of reference for the $Q_i\bar{Q}_i$ pair, the momenta of the individual quark and antiquark can be expressed in terms of the total momentum of the pair and their relative momentum. For a $Q_i(p_i)\bar{Q}_i(\bar{p}_i)$ pair, we denote $P_i$ as the total momentum and $q_i$ as the relative momentum of the pair. The momenta of the quark and antiquark are then expressed as 
\begin{eqnarray}
p_i &= \frac{1}{2} P_i + q_i, \
\bar{p}_i &= \frac{1}{2} P_i - q_i.
\label{eq.momentum1}
\end{eqnarray}
In the rest frame of the $Q_i\bar{Q}_i$ pair, the total momentum and relative momentum take the following forms: 
\begin{subequations}
\begin{eqnarray}
P_i  =&(2 E_i, 0),\\
q_i  =&(0, \bm{q}_i),\\
p_i  =&(E_i, \bm{q}_i), \\
\bar{p}_i  = &(E_i,-\bm{q}_i).
\end{eqnarray}
\end{subequations}
By definition, the total momentum and relative momentum are orthogonal, i.e., $P_i \cdot q_i = 0$. Applying the on-shell condition (with $m_{Q_i}$ being the mass of the heavy quark), the energy of each quark (and antiquark) is $E_i = \sqrt{m_{Q_i}^2 + \bm{q}_i^2}$, which gives rise to the following invariant quantities: 
\begin{equation}
p_i^{2} = \bar{p}_i^{2} = m_{Q_i}^{2}, \;
P_i^{2} = 4E_i^{2}, \;
q_i^{2} = -\bm{q}_i^{2} = -m_{Q_i}^{2} v_i^2.
\end{equation}

It is convenient to employ the covariant projection method to extract the contributions of different spin and color states. Spin projectors valid to all orders in the relative momentum have been derived in Refs. \cite{Bodwin:2002cfe,Bodwin:2010fi}. The corresponding projector for the spin-triplet states of heavy quark-antiquark pairs is
\begin{eqnarray}
\Pi_{3,i}&&=\frac{1}{\sqrt{2}(E_i+m_{Q_i})}\nonumber\\
&&\times(\slashed{\bar{p}_i}-m_{Q_i})\slashed{\epsilon}^*\frac{(\slashed{P_i}+2E_i)}{4E_i}(\slashed{p}_i+m_{Q_i}),
\end{eqnarray}
where $\epsilon$ denotes the polarization vector of the spin-triplet state. The projector for the $Q_i\bar{Q}_i$ color-singlet state takes the form 
\begin{equation}
\pi_{1}=\frac{1}{\sqrt{N_{c}}} \mathbf{1},
\end{equation}
where $\mathbf{1}$ denotes the unit color matrix. The $^3S_1^{[1]}$ contributions from the $Q_i\bar{Q}_i$ pairs are extracted by replacing the spinor products of $Q_i\left(\tfrac{P_i}{2} + q_i\right) \bar{Q}_i\left(\tfrac{P_i}{2} - q_i\right)$ in Eq.(\ref{eq.momentum1}) with the corresponding spin and color projectors for each quark flavor: $\Pi_{3,i}$ (for $i=1,2$) and $\pi_1$. The full amplitude is then obtained by evaluating the resulting spin and color traces after contraction with the projectors.

To expand the amplitude and examine the convergence of the series in $v$, we perform a Taylor expansion around $q_1 = q_2 = 0$, leading to: 
\begin{eqnarray}
&&{\cal M}_{Z \to Q_1\bar{Q}_1[\,^3S_1^{[1]}] + Q_2\bar{Q}_2[\,^3S_1^{[1]}]}\nonumber \\
&&= {\cal M}\Big{|}_{q_1=q_2=0}+\sum _{i=1,2}q_i^{\alpha} \frac{\partial {\cal M}}{\partial q_i^{\alpha}}\Big{|}_{q_1=q_2=0}\nonumber \\
&&~~~ +\sum _{i,j=1,2}\frac{1}{2!}q_i^{\alpha}q_j^{\beta}\frac{\partial^{2}{\cal M}}{\partial q_i^{\alpha} \partial q_i^{\beta}}\Big{|}_{q_1=q_2=0} +\cdots.
\label{eq.expand1}
\end{eqnarray}
To evaluate the amplitude according to Eq.(\ref{eq.expand1}), we obtain the relativistic corrections by expanding the S-wave amplitude in powers of ${\bm q}_i$ and averaging the ${\bm q}_i$-dependent tensors over the solid angles, where we use the following angular average relations: 
\begin{eqnarray}
\int \dfrac{d\Omega_{{\bm q}_i}}{4\pi}q_i^{\alpha} =0, ~ ~ ~
\int\!\dfrac{d\Omega_{{\bm q}_i}}{4\pi}q_i^{\alpha} q_j^{\beta} =\frac{\delta_{ij}{\bm q}_i^2}{3}I^{\alpha\beta},
\end{eqnarray}
where ${\bm q}_i^2 = |\bm q_i|^2 = m_{Q_i}^2 v_i^2$ (as defined earlier, since $q_i^2 = -\bm q_i^2 = -m_{Q_i}^2 v_i^2$), and
\begin{eqnarray}\label{eq_piexd}
I^{\alpha\beta}\equiv-g^{\alpha\beta}+\frac{P^{\alpha}P^{\beta}}{P^2}.
\end{eqnarray}
After expanding the amplitude up to relative order ${\cal O}(v^2)$ and averaging over the directions of $\bm q_i$, we obtain 
\begin{eqnarray}
&&{\cal M}_{Z \to Q_1\bar{Q}_1[\,^3S_1^{[1]}] + Q_2\bar{Q}_2[\,^3S_1^{[1]}]}\nonumber \\
&&= {\cal M}\Big{|}_{q_1=q_2=0} +\sum _{i}\frac{1}{2!}\frac{{\bm q}^2 I^{\alpha\beta}}{3}\frac{\partial^{2}{\cal M}}{\partial q_i^{\alpha} \partial q_i^{\beta}}\Big{|}_{q_1=q_2=0} +\cdots.
\label{eq.expand2}
\end{eqnarray}

According to Eqs. (\ref{eq.nrqcdfact-QQbar}) and (\ref{eq.expand2}), the SDCs at different orders can be expressed as: 
\begin{subequations}
\begin{eqnarray}
c_0^{(0)}&=&(\frac{{\cal M}^{(0)}}{\mathcal{N}_1 \mathcal{N}_2})\Big{|}_{q_1=q_2=0}\\
c_0^{(1)}&=&(\frac{{\cal M}^{(1)}}{\mathcal{N}_1 \mathcal{N}_2})\Big{|}_{q_1=q_2=0}\\
c_{2,1}^{(0)}&=&\frac{m_{Q_1}^2}{2!}\frac{ I_{1}^{\alpha\beta}}{3}\frac{\partial^{2}{\cal M}^{(0)}}{\partial q_1^{\alpha} \partial q_1^{\beta}\mathcal{N}_1 \mathcal{N}_2}\Big{|}_{q_1=q_2=0}\\
c_{2,2}^{(0)}&=&\frac{m_{Q_2}^2}{2!}\frac{ I_{2}^{\alpha\beta}}{3}\frac{\partial^{2}{\cal M}^{(0)}}{\partial q_2^{\alpha} \partial q_2^{\beta}\mathcal{N}_1 \mathcal{N}_2}\Big{|}_{q_1=q_2=0},
\end{eqnarray}
\end{subequations}
where the superscript ``$(i)$" of ${\cal M}^{(i)}$ indicates the order (in $\alpha_s$) of perturbation theory. The normalization factors $\mathcal{N}_i = 2\sqrt{6N_c} E_i$ arise from the LO LDMEs, and the squared LO LDMEs: 
\begin{eqnarray}
&& \vert \langle (Q_i\bar{Q}_i)^3S_1 \vert \psi_i^\dagger {{\bm \sigma}\cdot {\bm \epsilon}} \chi_i \vert 0 \rangle\vert^2=6N_c(2E_i)^2.
\end{eqnarray}

The SDCs $c_0^{(1)}$ is determined by matching the results obtained from the one-loop contributions to the amplitude ${\cal M}^{(1)}$. ${\cal M}^{(1)}$ receives contributions exclusively from virtual corrections. There are ultraviolet (UV) and infrared (IR) divergences in the one-loop diagrams. To handle these divergences, we employ dimensional regularization in $D = 4-2\epsilon$ dimensions to regularize these divergences. The UV divergences need to be removed through renormalization. We adopt on-shell renormalization conditions for both the heavy quark field ($Q_i$) and its mass ($m_{Q_i}$) in the heavy quark sector, yielding counterterms that contain both UV and IR poles:
\begin{eqnarray}
&&\delta Z_2^{\rm OS}=-\dfrac{C_F \alpha_s}{4\pi}\left[\dfrac{1}{\epsilon_{\rm UV}} +\dfrac{2}{\epsilon_{\rm IR}} -3\gamma_E +3\ln\dfrac{4\pi \mu_R^2}{m_{Q_i}^2} +4 \right], \nonumber \\
&&\delta Z_{m_i}^{\rm OS}=-\dfrac{3C_F \alpha_s}{4\pi}\left[\dfrac{1}{\epsilon_{\rm UV}} -\gamma_E +\ln\dfrac{4\pi \mu_R^2}{m_{Q_i}^2} +\frac{4}{3}\right],
\end{eqnarray}
where $C_F = 4/3$, $\gamma_E \approx 0.577$ is the Euler-Mascheroni constant, and $\mu_R$ is the renormalization scale. 

In dimensional regularization, special care must be taken in handling the $\gamma_5$ matrix. Throughout this work, we adopt the Larin prescription \cite{Larin:1993tq}, in which the product $\gamma_\mu\gamma_5$ is represented as
\begin{eqnarray}
\gamma_\mu\gamma_5=\frac{i}{3!}\epsilon_{\mu\rho\sigma\tau}\gamma^\rho\gamma^\sigma\gamma^\tau.
\end{eqnarray}
This scheme breaks the axial current Ward identity, and a finite renormalization of the axial vector vertex is required for its restoration. The corresponding renormalization constant $Z_5$ is given by
\begin{eqnarray}
Z_5=1-\frac{\alpha_s}{\pi}C_F.
\end{eqnarray}

In the calculation, FeynArts \cite{feynarts} is employed to generate Feynman diagrams. FeynCalc \cite{feyncalc1,feyncalc2} is used to evaluate the Dirac and color traces, while \$Apart \cite{apart} performs partial fractioning. The loop integrals are reduced to master integrals via integration-by-parts (IBP) reduction using FIRE \cite{fire}. Finally, these master integrals are evaluated numerically with LoopTools \cite{looptools}.

\section{Numerical results and discussions}
\label{sec4}
In the numerical calculation, the necessary input parameters are taken as follows: 
\begin{eqnarray}
&& m_c=1.4\pm 0.2 \,{\rm GeV},\; m_b=4.6 \pm 0.1\,{\rm GeV},\; \nonumber \\
&& m_{_Z}=91.1880\,{\rm GeV}, \;\, G_F=1.1663 \times 10^{-5}{\rm GeV}^{-2}, \nonumber \\
&& \alpha=1/128,
\label{eq.input}
\end{eqnarray}
where $m_c$ and $m_b$ denote the pole masses of the charm and bottom quarks, respectively, and their values are taken from Refs. \cite{Bodwin:2007fz,Chung:2010vz}. The values of $m_{_Z}$  and $ G_F$ are taken from the Particle Data Group (PDG) \cite{ParticleDataGroup:2024cfk}. For the running electromagnetic coupling constant, we adopt its value at the Z boson mass, $\alpha(m_Z) = 1/128$, which is used in the numerical calculation. For the running strong coupling constant $\alpha_s(\mu_R)$, we use the standard two-loop renormalization group equation (RGE): 
\begin{equation}
\alpha_s(\mu_R)
=\frac{4\pi}{\beta_0\, {\rm ln}(\mu_R^2/\Lambda_{\rm QCD}^2)}\left[ 1-\frac{\beta_1 {\rm ln} \,{\rm ln}(\mu_R^2/\Lambda_{\rm QCD}^2)}{\beta_0^2\, {\rm ln}(\mu_R^2/\Lambda_{\rm QCD}^2)} \right],
\label{eq.alphas}
\end{equation}
where $\beta_0 = 11 - 2n_f/3$, $\beta_1 = 102 - 38n_f/3$, and $n_f$ denotes the number of active light quark flavors. Using the input value $\alpha_s(m_Z) = 0.118$ \cite{ParticleDataGroup:2024cfk}, we determine the values of $\Lambda_{\rm QCD}^{n_f=5} = 0.226\ \text{GeV}$ and $\Lambda_{\rm QCD}^{n_f=4} = 0.328\ \text{GeV}$.

We take the color-singlet LDMEs for the vector quarkonium states $J/\psi$ and $\Upsilon(nS)$, defined as $\langle \mathcal{O}_{1}\rangle_{H_i} = \frac{1}{3}\sum_{\lambda}\vert \langle H_i(\lambda)\vert \psi^\dagger {\bm \sigma} \cdot {\bm \epsilon}(\lambda)\chi \vert 0 \rangle \vert^2$, from Refs. \cite{Bodwin:2007fz,Chung:2010vz}, i.e.,
\begin{subequations}
\begin{eqnarray}
&&\langle \mathcal{O}_{1}\rangle_{J/\psi}=0.440^{+0.067} _{-0.055}\,{\rm GeV}^{3}, \\
&&\langle \mathcal{O}_{1}\rangle_{\Upsilon(1S)}=3.069^{+0.207} _{-0.190} \,{\rm GeV}^{3},\\
&&\langle \mathcal{O}_{1}\rangle_{\Upsilon(2S)}=1.623^{+0.112} _{-0.103}\,{\rm GeV}^{3}, \\
&&\langle \mathcal{O}_{1}\rangle_{\Upsilon(3S)}=1.279^{+0.090} _{-0.083} \,{\rm GeV}^{3},
\end{eqnarray}
\label{eq.ldmes1}
\end{subequations}
and the matrix elements of the squared relative momentum $\langle \bm{q}^2\rangle_{H_i}$
\begin{subequations}
\begin{eqnarray}
&&\langle \bm{q}^2\rangle_{J/\psi}=0.441^{+0.140} _{-0.140}, \\
&&\langle \bm{q}^2\rangle_{\Upsilon(1S)}=-0.193^{+0.072}_{-0.073}, \\
&&\langle \bm{q}^2\rangle_{\Upsilon(2S)}=1.898^{+0.210}_{-0.210}, \\
&&\langle \bm{q}^2\rangle_{\Upsilon(3S)}=3.283^{+0.353}_{-0.352}.
\end{eqnarray}
\label{eq.ldmes2}
\end{subequations}
The quantities $\langle v^2\rangle_{H_i}$ are then derived directly using the relation $\langle v^2\rangle_{H_i} = \langle \bm{q}^2\rangle_{H_i}/m_{Q_i}^2$.

\subsection{Basic results}
In this subsection, we present the numerical results for the decay widths of $Z \to J/\psi+\Upsilon(nS)$ ($n=1,2,3$). For clarity, we first show the decay widths calculated using the central values of quark masses and LDMEs, followed by a systematic analysis of theoretical uncertainties from parameter variations.  

\begin{table}[htb]
	\begin{tabular}{c c c c c }
		\hline\hline
		$\alpha_s(\mu_R)$  &  ${\rm LO}$  & ${\cal O}(\alpha_s)$ & ${\cal O}(v^2)$  & {\rm Total} \\
		\hline
		$\alpha_s(2 m_b) =0.260$             &  $283.75$    &  $-194.42   $  & $-77.00$ & $ 12.32  $ \\
		$\alpha_s({m_{_Z}}/{2}) =0.132$  &  $283.75 $   &  $-140.66   $  & $-77.00$  & $ 66.09  $\\
		$\alpha_s(m_{_Z}) =0.118$             &  $283.75 $   &  $-125.82   $  & $-77.00$   & $ 80.93  $\\
		\hline\hline
	\end{tabular}
	\caption{Decay widths (in unit: $10^{-3}$eV) of $Z \to J/\psi+\Upsilon(1S) $ up to ${\cal O}(\alpha_{s})$ and ${\cal O}( v^2)$, with the contributions from different orders given explicitly.}
	\label{tb.1S}
\end{table}

\begin{table}[htb]
\begin{tabular}{c c c c c }
\hline\hline
$\alpha_s(\mu_R)$  &  ${\rm LO}$  & ${\cal O}(\alpha_s)$ & ${\cal O}(v^2)$  & {\rm Total} \\
\hline
$\alpha_s(2 m_b) =0.260$              &  $150.05$    &  $-102.81   $  & $-35.96$ & $ 11.28  $ \\
$\alpha_s({m_{_Z}}/{2}) =0.132$  &  $150.05 $   &  $-74.38   $  & $-35.96$  & $ 39.71  $\\
$\alpha_s(m_{_Z}) =0.118$             &  $150.05 $   &  $-66.54   $  & $-35.96$   & $ 47.56  $\\
\hline\hline
\end{tabular}
\caption{Decay widths (in unit: $10^{-3}$eV) of $Z \to J/\psi+\Upsilon(2S) $ up to ${\cal O}(\alpha_{s})$ and ${\cal O}( v^2)$, with the contributions from different orders given explicitly.}
\label{tb.2S}
\end{table}

\begin{table}[htb]
\begin{tabular}{c c c c c }
\hline\hline
$\alpha_s(\mu_R)$  &  ${\rm LO}$  & ${\cal O}(\alpha_s)$ & ${\cal O}(v^2)$  & {\rm Total} \\
\hline
$\alpha_s(2 m_b) =0.260$             &  $118.25$    &  $-81.02  $  & $-25.86$ & $ 11.37  $ \\
$\alpha_s({m_{_Z}}/{2}) =0.132$  &  $118.25 $   &  $-58.62   $  & $-25.86$  & $ 33.78  $\\
$\alpha_s(m_{_Z}) =0.118$             &  $118.25 $   &  $-52.43  $  & $-25.86$   & $ 40.00  $\\
\hline\hline
\end{tabular}
\caption{Decay widths (in unit: $10^{-3}$eV) of $Z \to J/\psi+\Upsilon(3S) $ up to ${\cal O}(\alpha_{s})$ and ${\cal O}( v^2)$, with the contributions from different orders given explicitly.}
\label{tb.3S}
\end{table}

The decay widths for $Z \to J/\psi+\Upsilon(nS)$ ($n=1,2,3$) are presented in Tables \ref{tb.1S}–\ref{tb.3S}, where the contributions from different orders are explicitly shown. These results are obtained using three typical renormalization scales: $2m_b$, $m_Z/2$, and $m_Z$. The tables report the leading-order (LO) decay widths, next-to-leading-order (NLO) QCD corrections at ${\cal O}(\alpha_s)$, and relativistic corrections at ${\cal O}(v^2)$ for the decays at different renormalization scales $\mu_R$ of $\alpha_s$, with all values given in units of $10^{-3}\ \text{eV}$.

\begin{figure}[htbp]
\includegraphics[width=0.4\textwidth]{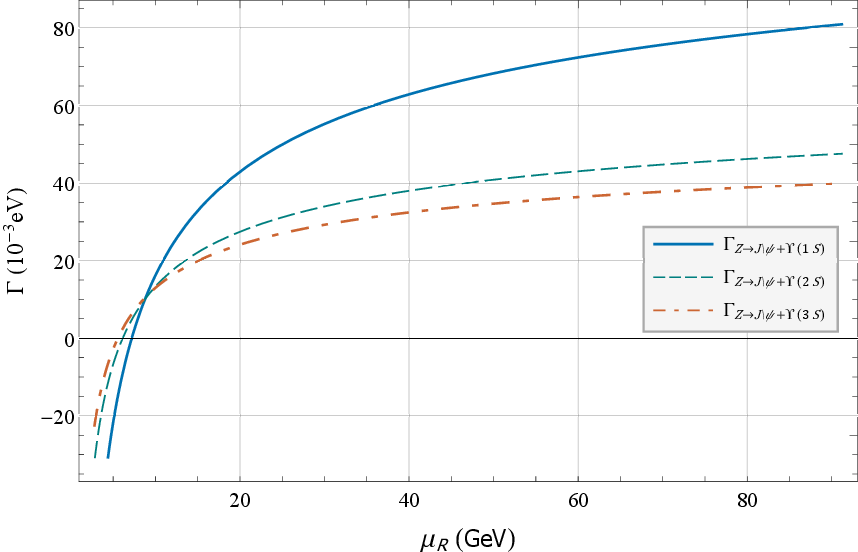}
\caption{Decay widths (in units of $10^{-3}\ \text{eV}$) of $Z \to J/\psi+\Upsilon(nS)$ ($n=1,2,3$) as a function of the renormalization scale $\mu_R$. The charm and bottom quark masses are fixed at $m_c=1.4\ \text{GeV}$ and $m_b=4.6\ \text{GeV}$, respectively, and $\mu_R$ is evolved from $2m_c$ to $m_Z$. $\Gamma$ denotes the total decay width, including contributions from LO, ${\cal O}(\alpha_s)$, and ${\cal O}(v^2)$.} \label{running_muR}
\end{figure}

The ${\cal O}(\alpha_s)$ QCD corrections and ${\cal O}(v^2)$ relativistic corrections make significant contributions to the decay widths of $Z \to J/\psi+\Upsilon(nS)$ decays. The LO decay widths are fixed for each $\Upsilon(nS)$ state, with central values of $0.284\ \text{eV}$, $0.150\ \text{eV}$, and $0.118\ \text{eV}$ for $\Upsilon(1S)$, $\Upsilon(2S)$, and $\Upsilon(3S)$, respectively. The ${\cal O}(\alpha_s)$ QCD corrections are strongly negative across all three renormalization scales, while the ${\cal O}(v^2)$ relativistic corrections take a constant negative value for each $\Upsilon(nS)$ state. At the central scale $\mu_R=m_Z/2$, the ${\cal O}(\alpha_s)$ corrections account for $-49.6\%$ of the LO decay width for all three $\Upsilon(nS)$ states, while the ${\cal O}(v^2)$ corrections account for $-27.1\%$, $-24.0\%$, and $-21.9\%$ for $\Upsilon(1S)$, $\Upsilon(2S)$, and $\Upsilon(3S)$, respectively. The ${\cal O}(\alpha_s)$ QCD correction fractions are identical across different $\Upsilon(nS)$ states, reflecting the fact that the NLO short-distance coefficient $c_0^{(1)}$ is universal for all $n$. In contrast, the ${\cal O}(v^2)$ correction fractions exhibit a moderate decrease with increasing radial quantum number $n$.

From Tables \ref{tb.1S}-\ref{tb.3S}, one can observe that the total decay widths become quite small at low renormalization scales, indicating that the fixed-order perturbative expansion becomes unreliable in this regime and that the inclusion of higher-order corrections is necessary to restore perturbative stability. To examine the dependence of the decay widths on the renormalization scale, we show this dependence in Fig. \ref{running_muR}. The curves show that the decay widths depend strongly on the scale of $\alpha_s$. 

\subsection{Uncertainty analysis}

In this subsection, we present a detailed estimation of the theoretical uncertainties for the decay widths of $Z \to J/\psi+\Upsilon(nS)$ ($n=1,2,3$). The primary sources of theoretical uncertainty arise from the choice of the renormalization scale $\mu_R$, the heavy quark masses ($m_c$ and $m_b$), the color-singlet LDMEs $\langle \mathcal{O}_1 \rangle_{H_i}$, and the matrix elements of the squared relative momentum $\langle \bm{q}^2 \rangle_{H_i}$. 

Strictly speaking, the heavy-quark masses, the LDMEs, and $\langle \bm{q}^2 \rangle_{H_i}$ are not completely independent, as they are extracted within a unified fitting framework~\cite{Bodwin:2007fz,Chung:2010vz}. To avoid double counting and partially capture their correlations, we decompose the total uncertainties of the LDMEs provided in Ref.~\cite{Bodwin:2007fz} into their individual origins. Specifically, the variations in the LDMEs induced by the heavy-quark masses and $\langle \bm{q}^2 \rangle_{H_i}$ are isolated and evaluated synchronously when varying $m_Q$ and $\langle \bm{q}^2 \rangle_{H_i}$, respectively. The remaining components of the LDME uncertainties (e.g., those arising from the string tension, estimated NNLO corrections, and $\alpha_s$) are combined in quadrature and assigned as the independent intrinsic LDME uncertainty. As will be shown below, any residual correlation effects under this approximation are heavily overshadowed by the renormalization-scale uncertainty, which dominates the total error budget.

In our practical evaluation, we estimate the impact of each parameter individually while fixing the others at their central values. The upper and lower deviations from each error source are calculated separately, and the total upper (lower) theoretical uncertainty is obtained by combining the corresponding individual components in quadrature. The specific variation ranges and strategies are set as follows:
(1) Renormalization scale: We vary $\mu_R$ between $2m_b$ and $m_Z$, with $m_Z/2$ as the central scale.\footnote{It is worth noting that at NLO accuracy, the explicit $\ln(\mu_R^2)$ dependence from the loop integrals is exactly canceled by that from the counterterms once all divergences (including both ultraviolet and infrared poles) are removed. Consequently, the renormalization scale dependence of the decay width stems solely from the running coupling $\alpha_s(\mu_R)$. Given the multiple well-separated physical scales involved in this process (ranging from the heavy quark masses up to $m_{_Z}$), there is no unambiguously optimal choice for $\mu_R$ at the one-loop level. Therefore, to ensure a conservative estimation of the theoretical uncertainty, we vary $\mu_R$ over a broad range from $2m_b$ to $m_{_Z}$, leaving a more rigorous scale-setting (e.g., using the Principle of Maximum Conformality~\cite{Brodsky:2011ta,Brodsky:2011ig,Brodsky:2012rj,Mojaza:2012mf,Brodsky:2013vpa}) for future higher-order studies.}
(2) Heavy quark masses: We vary $m_c=1.4\pm 0.2\ \text{GeV}$ and $m_b=4.6\pm 0.1\ \text{GeV}$ around their central values.
(3) Relativistic matrix elements: $\langle \bm{q}^2 \rangle_{H_i}$ are varied within the ranges given in Eq.~(\ref{eq.ldmes2}).
(4) LDMEs: $\langle \mathcal{O}_1 \rangle_{H_i}$ are varied according to their independent intrinsic uncertainties (after removing the $m_Q$ and $\langle \bm{q}^2 \rangle_{H_i}$ induced errors, as discussed above).

The individual uncertainties evaluated from these sources are listed as follows:
\begin{subequations}
\begin{eqnarray}
&&\Gamma_{Z\to J/\psi+\Upsilon(1S)}\nonumber\\&&=66.09_{-53.76-20.63-9.02-22.80}^{+14.83+3.48+10.80+22.97}\,10^{-3}\,{\rm eV}, \label{Za1}\\
&&\Gamma_{Z\to J/\psi+\Upsilon(2S)}\nonumber\\&&=39.71_{-28.43-8.10-5.45-12.06}^{+7.85+1.98+6.51+12.15}\,10^{-3}\,{\rm eV}, \label{Zb1}\\
&&\Gamma_{Z\to J/\psi+\Upsilon(3S)}\nonumber\\&&=33.78_{-22.40-4.95-10.16-9.50}^{+6.18+1.20+10.59+9.57}\,10^{-3}\,{\rm eV}, \label{Zc1}
\end{eqnarray}
\end{subequations}
where the upper uncertainties (superscripts) and lower uncertainties (subscripts) correspond to the renormalization scale, heavy quark masses, $\langle \mathcal{O}_1 \rangle_{H_i}$, and $\langle \bm{q}^2 \rangle_{H_i}$ (combined LDME uncertainties), in sequence. We then combine these individual uncertainties in quadrature to obtain the total theoretical uncertainties for the decay widths, given by
\begin{subequations}
\begin{eqnarray}
&&\Gamma_{Z\to J/\psi+\Upsilon(1S)}=66.09 ^{+29.61} _{-62.59} \,\, 10^{-3}{\rm eV}, \label{Za2}\\
&&\Gamma_{Z\to J/\psi+\Upsilon(2S)}=39.71 ^{+15.98 } _{-32.39 } \,\, 10^{-3}{\rm eV}, \label{Zb2}\\
&&\Gamma_{Z\to J/\psi+\Upsilon(3S)}=33.78^{+15.60} _{-26.83 } \,\, 10^{-3}{\rm eV}.\label{Zc2}
\end{eqnarray}
\end{subequations}

Using the predicted decay widths from this work and the total $Z$-boson decay width $\Gamma_Z=2.4955\ \text{GeV}$ from the Particle Data Group (PDG) \cite{ParticleDataGroup:2024cfk}, we calculate the branching fractions for $Z\to J/\psi+\Upsilon(nS)$ ($n=1,2,3$), with all branching fractions given in:
\begin{subequations}
\begin{eqnarray}
{\rm Br}(Z\to J/\psi+\Upsilon(1S))&=&26.48 ^{+11.86} _{-25.08}\times 10^{-12},\label{Za3}\\
{\rm Br}(Z\to J/\psi+\Upsilon(2S))&=&15.91 ^{+6.40} _{-12.98}\times 10^{-12},\label{Zb3}\\
{\rm Br}(Z\to J/\psi+\Upsilon(3S))&=&13.53 ^{+6.25} _{-10.75}\times 10^{-12}.\label{Zc3}
\label{eq.Br.psi}
\end{eqnarray}
\end{subequations}

\section{Summary}
\label{sec5}

We have presented a complete calculation of the decay widths for the rare Z-boson decays $Z\rightarrow J/\psi+\Upsilon(nS)$ ($n=1,2,3$) within the NRQCD framework, including NLO QCD corrections at $\mathcal{O}(\alpha_s)$ and relativistic corrections at $\mathcal{O}(v^2)$. Our numerical analysis reveals that both higher-order corrections are sizable and negative, severely suppressing the leading-order decay widths. Despite this suppression, the absolute magnitudes of the individual $\mathcal{O}(\alpha_s)$ and $\mathcal{O}(v^2)$ corrections remain strictly smaller than the LO contribution at the central scale (${m_{_Z}}/{2}$), suggesting that the perturbative expansions in $\alpha_s$ and $v$ maintain reasonable behavior.  

The observed reduction in decay widths is largely attributed to significant numerical cancellations between the LO term and the higher-order corrections, which subsequently lead to substantial total theoretical uncertainties. The predicted branching fractions are ${\rm Br}(Z\to J/\psi+\Upsilon(1S))=26.48 ^{+11.86} _{-25.08}\times 10^{-12}, {\rm Br}(Z\to J/\psi+\Upsilon(2S))=15.91 ^{+6.40} _{-12.98}\times 10^{-12}, {\rm Br}(Z\to J/\psi+\Upsilon(3S))=13.53 ^{+6.25} _{-10.75}\times 10^{-12}$.  

As discussed earlier, future high-luminosity $e^+e^-$ facilities (such as the ILC, CEPC, FCC-ee, and Super Z factory) are expected to produce an enormous number of Z-boson events. Based on the projected sample of $5 \times 10^{12}$ Z bosons at the FCC-ee \cite{FCC:2018evy}, an order-of-magnitude estimate using our central values suggests that approximately 130 $Z\rightarrow J/\psi+\Upsilon(1S)$ events could be produced. Given the current perturbative instability, this should be viewed as an initial theoretical baseline rather than a precise phenomenological benchmark. Our study demonstrates that higher-order corrections in these channels are too significant to neglect, highlighting the need to consider even higher-order corrections or resum large logarithms to improve the accuracy of theoretical predictions.  

\hspace{2cm}

\noindent {\bf Acknowledgments:} This work was supported in part by the Natural Science Foundation of China under Grants No. 12447183 and No. 12547101.

\hspace{2cm}

\end{document}